\documentclass[amsmath,amssymb,prr,aps,showpieces,superscriptaddress,twocolumn,bibnotes,floatfix]{revtex4-2}
\usepackage{amsmath,amsfonts,amssymb,amsthm,graphics,graphicx,epsfig,bbm}
\usepackage[colorlinks=true,citecolor=blue,linkcolor=blue,urlcolor=blue]{hyperref}
\usepackage[usenames]{color}
\usepackage{graphicx}
\usepackage{subfigure}
\usepackage{amsmath}
\usepackage{epsfig}
\usepackage{dcolumn}
\usepackage{bm}
\usepackage{color}
\usepackage{times}
\usepackage{epstopdf}
\usepackage[dvipsnames]{xcolor}
\usepackage{amssymb}
\usepackage{amstext}
\usepackage{latexsym}
\usepackage{comment}
\usepackage{hyperref}
\usepackage{amsfonts}
\usepackage{psfrag}
\usepackage{soul,xcolor}
\usepackage[normalem]{ulem}
\usepackage{dsfont}
\usepackage{txfonts}
\usepackage{float}
\usepackage{algorithm}
\usepackage{algpseudocode}
\usepackage{booktabs}
\usepackage{dblfloatfix}
\usepackage{caption}
\usepackage[percent]{overpic}

\newcommand{\dg}[1]{{\color{blue}(DG: \textit{#1})}} 

\usepackage{xcolor}

\newcommand{\hd}[1]{{\color{green!60!black}(HD: \textit{#1})}}

\newcommand{\ket}[1]{\vert #1 \rangle}
\newcommand{\bra}[1]{\langle #1 \vert}

\newcommand{\Tr}{\mathrm{Tr}}
\newcommand{\etal}{{\it{et al.}}}

\usepackage{tikz,xcolor,hyperref}
\definecolor{lime}{HTML}{A6CE39}
\DeclareRobustCommand{\orcidicon}{%
	\begin{tikzpicture}
	\draw[lime, fill=lime] (0,0) 
	circle [radius=0.16] 
	node[white] {{\fontfamily{qag}\selectfont \tiny ID}};
	\draw[white, fill=white] (-0.0625,0.095) 
	circle [radius=0.007];
	\end{tikzpicture}
	\hspace{-2mm}
}

\usepackage{orcidlink} 

\foreach \x in {A, ..., Z}{%
	\expandafter\xdef\csname orcid\x\endcsname{\noexpand\href{https://orcid.org/\csname orcidauthor\x\endcsname}{\noexpand\orcidicon}}
}

\renewcommand{\orcidB}[1]{\orcidlink{#1}}

\begin{document}

\setstcolor{red}

\title{Counterdiabatic ADAPT-VQE for molecular simulation}
\date{\today}

\author{D. Tancara \orcidB{0000-0002-5053-3521}} 
\affiliation{Facultad de Física, Pontificia Universidad Católica de Chile, Santiago 7820436, Chile}

\author{H. Díaz-Moraga \orcidB{0009-0007-1446-9060}}
\affiliation{Facultad de Física, Pontificia Universidad Católica de Chile, Santiago 7820436, Chile}

\author{D. Goyeneche
\orcidB{0000-0002-9865-4226}}
\affiliation{Facultad de Física, Pontificia Universidad Católica de Chile, Santiago 7820436, Chile}

\begin{abstract}    
Among variational quantum algorithms designed for NISQ devices, ADAPT-VQE stands out for its robustness against barren plateaus, particularly in estimating molecular ground states. On the other hand, counterdiabatic algorithms have shown advantages in both performance and circuit depth when compared to standard adiabatic approaches. In this work, we propose a hybrid method that integrates the ADAPT-VQE framework with counterdiabatic driving within an adiabatic evolution scheme. Specifically, we map the molecular Hamiltonian to a qubit representation and construct an adiabatic Hamiltonian, from which an approximate adiabatic gauge potential is computed using nested commutators. The resulting operator terms define the operator pool, and the ADAPT-VQE algorithm is applied to iteratively select the most relevant elements for the ansatz. Our results demonstrate improvements in performance and reductions in circuit depth compared to using either counterdiabatic algorithms or ADAPT-VQE with fermionic excitation operators, thus supporting the effectiveness of combining both paradigms in molecular simulations.
\end{abstract}
\maketitle

\section{Introduction}

Research in quantum computing has expanded significantly over the last few decades, with several experimental demonstrations of quantum advantage emerging even under the constraints of current noisy intermediate-scale quantum (NISQ) devices \cite{AruteNature2019, Zhong2021PhysRevLett, Wu2021PhysRevLett, Madsen2022Nature, Kim2023Nature, Zhu2022SciBull, Morvan2024Nature, Acharya2024Nature, Gao2025PhysRevLett}. This has motivated the development of quantum algorithms specifically designed for the NISQ era, which has become an active area of research \cite{BhartiRevModPhys2022}. In this context, the variational quantum eigensolver (VQE) has emerged as a promising alternative, combining a shallow quantum circuit built from a chosen ansatz with classical optimization to approximate ground states efficiently on NISQ devices. In the foundational work \cite{PeruzzoNatCommun2014}, an ansatz for molecular simulations was proposed based on the unitary coupled cluster with single and double excitations (UCCSD). This approach involves the exponential of a linear combination of fermionic excitation operators, where the variational parameters correspond to the weights assigned to each excitation term. Another notable approach apply to molecular simulations is the hardware-efficient ansatz \cite{KandalaNature2017}, in which local operations are implemented using standard quantum gates, and entanglement between qubits is generated through direct hardware-level interactions. 

Across the different ansatz employed in VQE and in variational quantum algorithms in general, the barren plateau problem has emerged as a serious challenge and has been extensively investigated in recent years \cite{FontanaNatCommun2024, RagoneNatCommun2024}. This problem is reflected in an exponential suppression of the cost-function gradient variance as the system size increases, leading to severe training difficulties for variational quantum algorithms when scaling to larger numbers of qubits. This can be characterized by the exponential vanishing variance of the cost function gradients with respect to the system size, which makes variational quantum algorithms increasingly difficult to train as the number of qubits grows. Consequently, designing algorithms that incorporate reliable mechanisms to alleviate barren plateaus continues to be an active and relevant line of research.
Among the limited set of variational quantum algorithms that are robust to barren plateaus, ADAPT-VQE stands out as a notable example \cite{GrimsleynpjQuantumInf2023}, with its principal characteristic being the dynamic construction of the ansatz \cite{GrimsleyNatCommun2019}. Starting from a reference state, the circuit is iteratively constructed by selecting and appending parameterized unitaries from a predefined operator pool. This selection process is guided by the energy gradient associated with each operator, ensuring that the ansatz evolves along the steepest descent direction to progressively drive the system toward its ground state. 
Initially, ADAPT-VQE was introduced for molecular simulations, where the operator pool is composed of single and double excitations operators mapped to the qubit operators. These operators, which correspond to those used in the exponential form of the UCCSD ansatz, generates the operator pool. By adaptively selecting only the most relevant components, the algorithm not only achieves higher accuracy but also leads to significantly shallower circuits, ultimately outperforming standard UCCSD ansatz. Subsequently, alternative operator pools were introduced, in which the fermionic single and double excitations are replaced by their qubit operators representation through direct Pauli strings \cite{LunTangPRXQuantum2021}, leading to an improvement in circuit depth compared with the original ADAPT-VQE formulation. Moreover, incorporating qubit excitations directly, rather than relying on fermionic excitations mapped to the qubit representation \cite{YordanovCommunPhys2021}, and including double qubit excitations within the same spin-orbitals \cite{RamoanpjQuantumInf2025}, has led to improved circuit depth and enhanced accuracy.

The ADAPT-VQE has also been applied to problems in another contexts. For instance, an alternative version of the algorithm was employed to prepare the 100-qubit vacuum state of the Schwinger model using a superconducting qubit quantum computer \cite{FarrellPRXQuantum2024}. A modified formulation of ADAPT-VQE was also proposed for quantum dynamics simulation, incorporating McLachlan’s variational principle in the cost function \cite{YaoPRXQuantum2021}. These results highlight the versatility of the ADAPT-VQE strategy and have motivated further developments, including improved measurement protocols and alternative adaptive schemes \cite{AnastasiouPRR2024, VaqueroSabaterJCTC2025}.

Another approach for ground state preparation is digitized counterdiabatic quantum optimization (DCQO) \cite{HegadePRResearch2022}. This method can be understood as a digitized implementation of adiabatic quantum computing enhanced with counterdiabatic driving that are introduced to suppress nonadiabatic transitions during the evolution. In contrast to standard adiabatic protocols, which require long evolution times to follow the adiabatic theorem, counterdiabatic driving enables the system to closely follow the instantaneous ground state even for finite-time evolutions. The digitization of the protocol through Trotterization of the time-evolution operator makes DCQO compatible with gate-based quantum hardware, enabling the preparation of ground states of optimization and many-body Hamiltonians with reduced circuit depth compared to digitized adiabatic evolution \cite{HegadePRResearch2022}. The DCQO has been  applied to different problems like protein folding \cite{ChandaranaPRAppl2023}, logistic scheduling \cite{DalalPRAppl2024}, portfolio optimization \cite{CadavidPhysRevApplied2024} and recently to molecular simulation \cite{FerreiroVelezArXiv2024}.

In this work, we propose a hybrid quantum–classical algorithm that combines the ADAPT-VQE strategy for operator selection and ansatz construction with concepts from counterdiabatic quantum computing to approximate ground states. Focusing on the shorter evolution times, we retain only the counterdiabatic driving terms and use them to define an operator pool for the ADAPT-VQE implementation. This approach is applied to the ground state estimation of molecular Hamiltonians. 

\section{Methods}
In this section, we outline the construction of the adiabatic Hamiltonian starting from a molecular Hamiltonian, review the standard counterdiabatic protocol, and finally present our algorithm.

\subsection{Molecular Hamiltonian to adiabatic Hamiltonian}

The determination of molecular ground states and energies is fundamentally governed by the electronic structure, which are the coulomb electronic interactions between electrons and nuclei. In the low-energy regime, the Born-Oppenheimer approximation \cite{McArdleRevModPhys2020} is well-justified by the fact that nuclei are significantly heavier than electrons. This approximation treats nuclei as stationary point charges, effectively decoupling nuclear and electronic motion. Within this approximation, the problem reduces to solving the time-independent Schrödinger equation for the electronic Hamiltonian, $H_f$, given by: 
\begin{eqnarray}\label{eq1: 1st_quant_elec_struc_ham}
    H_f =  - \frac{1}{2}  \sum \limits_i \nabla^2_{r_i}
    - \sum \limits_i \sum \limits_j \frac{Z_j}{|\mathbf{R}_j - \mathbf{r}_i|} 
     + \sum \limits_{i < j} \frac{1}{|\mathbf{r}_i - \mathbf{r}_j|},
\end{eqnarray}
where $\mathbf{r}_i$ and $\mathbf{R}_j$ denote the positions of electrons and nuclei, respectively, and $Z_j$ is the atomic number of the $j$-th nucleus.
This Hamiltonian encapsulates the kinetic energy of the electrons, the Coulombic attraction between electrons and nuclei, and the inter-electronic repulsion in atomic units (au).

To encode this problem into a quantum computer, we follow the standard second quantization formalism, where the Hamiltonian is written in terms of fermionic creation and annihilation operators acting on Fock space:
\begin{eqnarray}
        \hat{H}_f = \sum_{ij} h_{ij} \hat{a}^\dagger_j \hat{a}_i 
    + \frac{1}{2} \sum_{ijkl} h_{ijkl} \hat{a}^\dagger_l \hat{a}^\dagger_k \hat{a}_j \hat{a}_i.
\end{eqnarray}
Here, the coefficients \( h_{ij} \) and \( h_{ijkl} \) are the one-body and two-body integrals, respectively, defined as:
\begin{eqnarray}
    h_{ij} = \int \phi_i^*(\mathbf{r}) \left( -\frac{1}{2} \nabla^2 - \sum \limits_n \frac{Z_n}{r_{n}} \right) \phi_j(\mathbf{r})\, d\mathbf{r}, \\
    h_{ijkl} = \iint \frac{\phi_i^*(\mathbf{r}_1) \phi_j^*(\mathbf{r}_2) \phi_k(\mathbf{r}_2) \phi_l(\mathbf{r}_1)}{|\mathbf{r}_1 - \mathbf{r}_2|} \, d\mathbf{r}_1 d\mathbf{r}_2.
\end{eqnarray}

Where the spin-orbital functions $\{ \phi_k \}$ are the set of basis which the fock space is represented. Chemical basis are utilized in order to express approximately the molecular orbitals of the molecules.

Finally, the isomorphism between the fermionic algebra and the $n$ qubit algebra $\mathfrak{su}(2^n)$ is established via mappings such as the Jordan-Wigner, Parity, or Bravyi-Kitaev transformations \cite{SeeleyJCP2012}. This allows the Hamiltonian to be fully expressed in terms of Pauli strings:
\begin{eqnarray}\label{eq:H_es_Pauli}
    \hat{H}_f =\sum \limits_{\hat{P}_i \in \mathcal{P}} h_i \hat{P}_i
\end{eqnarray}
where $h_i$ is the coefficient associated to the $P_i$ Pauli string acting on the space of $n$ qubits, and $\mathcal{P} = {\{ \hat{I},\hat{X},\hat{Y},\hat{Z} \} ^{\otimes n}}$ the Pauli group associated to it.

Adiabatic quantum computing exploits the adiabatic theorem in order to compute the ground state of a target Hamiltonian, $\hat{H}_f$, by preparing a known ground state of an initial Hamiltonian $\hat{H}_i$ and by taking into account the evolution of a schedule function $\lambda(t)$ in the following adiabatic Hamiltonian:
\begin{eqnarray}
    \hat{H}_{\text{ad}} = \left( 1 - \lambda(t) \right) \hat{H}_i + \lambda(t) \hat{H}_f,
\end{eqnarray}where schedule function satisfies $\lambda(0)=0$ and $\lambda(T) = 1$, with $T$ the final time, therefore the adiabatic Hamiltonian ends in $\hat{H}_f$.
While the final Hamiltonian is determined by the specific problem to be solved—in this case, the molecular Hamiltonian $H_f$ expressed as a linear combination of Pauli strings \ref{eq:H_es_Pauli}—the initial Hamiltonian $H_i$ is selected for its efficiency in enabling easy ground-state preparation. As established in the literature \cite{McArdleRevModPhys2020}, for molecular problems, the Hartree-Fock state $\ket{\text{HF}}$ is a good initial reference state. In the second quantization framework, the Hartree-Fock state is defined as the single Slatter determinant of spin-orbitals. By defining the occupation vector $\mathbf{o} = (o_{n-1}, \dots, o_0)$, where $o_k = 1$ for the $N$ lowest-energy spin-orbitals and $o_k = 0$ otherwise. Depending on the fermion-to-qubit mapping utilized this occupation vector is transformed into a specific computational basis state: $\ket{\text{HF}} = \ket{b_{n-1}b_{n-2}, \dots, b_{0} }$, where the bitstring $\{b_k\}$ is uniquely determined by the mapping and the number of electrons in the basis. To follow the adiabatic protocol, the initial state must be the ground state of the initial Hamiltonian $\hat{H}_i$. We therefore define the local initial Hamiltonian $\hat{H}_i$ for encoding the Hartree-Fock state as the ground state:
\begin{eqnarray}
    \hat{H}_i = \sum_{k= 0}^{n-1} (-1)^{b_{k} + 1} \hat{Z}_k \label{Hi}
\end{eqnarray}
where $\hat{Z}_k \equiv \mathbb{I}^{\otimes n-k}  \otimes \hat{Z} \otimes \mathbb{I}^{\otimes k-1}$ is the local $\hat{Z}$ Pauli gate acting on the $k$-th qubit and $\mathbb{I}$ is the local identity operator. This initial Hamiltonian is chosen because it is easy to implement due to the locality of its operators, and because its ground state corresponds to the Hartree–Fock state when this state is known. To illustrate this, we consider a simple case of four spin-orbitals represented by four qubits for a two-electron molecule. Following the ordering $\ket{\alpha_1\alpha_0\beta_1\beta_0}$, where $\alpha_i$ ($\beta_i$) represents the occupational state of the $i-$th spin up (down) spatial orbital, the Hartree–Fock state is $\ket{\mathrm{HF}} = \ket{0101}$. The initial Hamiltonian is then given by:
\begin{eqnarray}
    \hat{H}_i = -\hat{Z}_0 + \hat{Z}_1 - \hat{Z}_2 + \hat{Z}_3,
\end{eqnarray}
where $\hat{Z}\ket{0}=\ket{0}$ and $\hat{Z}\ket{1}=-\ket{1}$. We can verify that $\ket{0101}$ is the ground state with eigenvalue $-4$. Therefore, using Eq.~\ref{Hi} we can encode the $n$-qubit Hartree-Fock state as the ground state and construct the corresponding adiabatic Hamiltonian.

\subsection{Counterdiabatic protocol}

Since adiabatic evolution must proceed slowly in order to satisfy the adiabatic theorem, its direct use can become impractical for quantum computing. To mitigate this limitation, one can incorporate an additional counterdiabatic driving term that accelerates the protocol and suppresses diabatic excitations. The total Hamiltonian with counterdiabatic driving is:
\begin{eqnarray} \label{eq:counterdiab_ham}
        \hat{H}_{\text{cd}} = \left( 1 - \lambda(t) \right) \hat{H}_i + \lambda(t) \hat{H}_f + \dot{\lambda}(t) \hat{A}_\lambda,
\end{eqnarray}
where $\dot{\lambda}(0)= \dot{\lambda}(T)=0$ and $\hat{A}_\lambda$ is the adiabatic gauge potential (AGP). In practice, evaluating $\hat{A}_\lambda$ requires access to the instantaneous eigenstates and eigenvalues of the adiabatic Hamiltonian $\hat{H}_{\text{ad}}$, which makes its direct computation impractical. However, we can calculate approximate versions of the AGP without the spectral information of $\hat{H}_{\text{ad}}$.
One of such approximation is given in Ref. \cite{ClaeysPRL2019} and is based in an expansion of nested commutators. The corresponding $l$-th order approximation can then be written as:
\begin{eqnarray}
\hat{A}_{\lambda}^{(l)} = i \sum_{k=1}^{l} \alpha_k(t) \hat{O}_{2k-1}(t)
\end{eqnarray}
where the nested commutators are:
\begin{eqnarray}
    \hat{O}_k = \underbrace{[\hat{H}_{\textrm{ad}},[\hat{H}_{\textrm{ad}},...[\hat{H}_{\textrm{ad}}}_{k},\partial_\lambda \hat{H}_{\textrm{ad}}]]],
\end{eqnarray}
and the coefficients $\alpha_k(t)$ can be obtained by minimizing the action $S_l = \Tr{[G_l^2]}$, with $G_l =\partial_\lambda \hat{H}_{\textrm{ad}}- i[\hat{H}_{\text{ad}}, \hat{A}_{\lambda}^{(l)} ]$. In the limit $l \to \infty$, the series reproduces the exact AGP. The counterdiabatic protocol has been used in the context of quantum computing with an approximate AGP, typically with $l = 1$, together with a Trotter-Suzuki decomposition to digitize the time-evolution operator generated by $\hat{H}_{\text{cd}}$, showing improvements in circuit depth and performance with respect to the standard adiabatic protocol \cite{HegadePRResearch2022}.

\subsection{Counterdiabatic ADAPT-VQE}

The VQE has become one of the central algorithms in the NISQ era, particularly in quantum chemistry. However, its accuracy depends greatly on the choice of ansatz, which has motivated the development of many tailored constructions designed to balance expressiveness and circuit depth. With the parametrized ansatz, the state $\ket{\psi(\boldsymbol{\theta})}$ is used in a classical optimization to minimize the energy, which is given by
\begin{eqnarray}
    E = \min_{\boldsymbol{\theta}}\bra{\psi(\boldsymbol{\theta})}\hat{H}_f\ket{\psi(\boldsymbol{\theta})}
\end{eqnarray}
Common fixed ansatz for molecular simulations are based on excitation operators, which often lead to large circuit depths that are susceptible to barren plateaus.
In this context, ADAPT-VQE algorithm was introduced to addresses these issues by building the ansatz iteratively from energy gradient criterion, yielding more compact circuits and showing improved robustness against barren plateaus, thus enhancing its potential for near-term quantum applications. 

The ADAPT-VQE algorithm begins by defining an operator pool consisting of a set of operators $\{\hat{V}_k\}_{k=1}^M$, where the original choice for $\hat{V}_k$ corresponds to the single and double excitation operators mapped to the qubit representation. Their Pauli decomposition leads to the Pauli strings used in more recent formulations, so that in general $\hat{V}_k \in \{\hat{X}, \hat{Y}, \hat{Z}, \hat{I}\}^{\otimes n}$.
The algorithm is initialized with a reference state $\ket{\psi_i}$ at iteration $j = 0$, usually the Hartree-Fock state. For each operator $\hat{V}_k$ in the pool, the energy gradient with respect to a variational parameter $\theta_k$ is evaluated when applying the unitary $e^{-i\theta_k \hat{V}_k}$ to the current ansatz state $\ket{\psi_j}$, at $\theta_k = 0$. This derivative takes the form of a commutator expectation value:

\begin{eqnarray}
g_k = \left. \frac{\partial E}{\partial \theta_k} \right|_{\theta_k = 0} = i\bra{\psi_j} [\hat{V}_k,\hat{H}] \ket{\psi_j},
\end{eqnarray}
 and the collection of all such values defines the gradient vector $\vec{g} = (g_1, \dots, g_M)$. If the 2-norm $\|\vec{g}\|_2$ is less than a predefined threshold $\epsilon$, the algorithm terminates. Otherwise, the operator $V_k$ corresponding to the maximum absolute component of $\vec{g}$ is selected (denoted $V^{\text{max}}$) and added to the ansatz. The ansatz is then updated as:

\begin{eqnarray}
\ket{\psi_{j+1}} = e^{-i\theta \hat{V}^{\text{max}}} \ket{\psi_j},
\end{eqnarray}
where $\theta$ is a newly introduced variational parameter. A VQE optimization is then performed to update all parameters to the optimum value and obtain the new optimal state $\ket{\psi_{j+1}^{\text{opt}}}$. This state is then set as the input for the next iteration:\mbox{
$\ket{\psi_{j+1}^{\text{opt}}} \rightarrow \ket{\psi_j}$} and the process repeats until $\|\vec{g}\|<\epsilon$.

The choice of an appropriate operator pool is a crucial step in the ADAPT-VQE algorithm. In this work we propose constructing the operator pool using the approximate AGP as follows. The counterdiabatic evolution can be analyzed in different regimes determined by the total evolution time $T$. For sufficiently short final times, the Hamiltonian varies rapidly, making the counterdiabatic contribution more relevant than the adiabatic term. This situation is commonly referred to as the impulse regime \cite{CadavidPhysRevApplied2024}, in which the dynamics is dominated by the rate of change of the schedule function and satisfies the condition $|\lambda(t)| \ll |\dot{\lambda}(t)|$. Under this assumption, the Hamiltonian can be approximated by:
\begin{eqnarray}
    \hat{H}_{\text{cd}} \approx \dot{\lambda}(t)\hat{A}_{\lambda}^{(l)}.\label{cd_hamiltonian}
    \end{eqnarray} Once performed the mapping that transforms $\hat{H}_{\text{cd}}$ to the Pauli set, the $l$-th order approximated AGP can be writen in terms of generators or Pauli strings by solving the nested commutators:
\begin{eqnarray}
    \hat{A}_{\lambda}^{(l)} = \sum_{j= 1}^{\eta} a_j(t) \hat{G}_j
\end{eqnarray}where $\eta$ is the number of Pauli strings of $\hat{A}_{\lambda}^{(l)}$ and the coefficients ${a}_i(t)$ are weights associated to the $j$-th Pauli strings $G_j$, that are obtained by the ponderation of the terms $\alpha_k(t)$, and combinations of powers of $\lambda(t)$ and $\left( 1-\lambda(t) \right)$ arising from the nested commutators. The operator pool is then constructed by fixing the $l$-th order giving rise to the operator pool $\{ G_j \}_{j=1}^{\eta}$. 

This operator pool is motivated by the dynamics of the counterdiabatic protocol, and in particular by the Trotterized time-evolution operator generated by the counterdiabatic Hamiltonian in Eq.~\ref{cd_hamiltonian}, which takes the form:
\begin{eqnarray}\label{eq:trotterized_U(T)_at_final_time}
    \hat{U}(T) = \prod_{k=1}^{p} \prod_{j=1}^{\eta}
\exp\!\left[-i\,\dot{\lambda}(k \delta t)a_j(k \delta t)\,\delta t\, \hat{G}_j \right].
\end{eqnarray}
The expression above has been implemented on quantum processors as a digitized counterdiabatic quantum algorithm, where in practice only the case $l = 1$ is used, since $\eta$ increases significantly with the $l$-th AGP approximation, increasing the circuit depth. Related works have also used the time-evolution operator of the counterdiabatic protocol as inspiration to construct a variational ansatz, replacing the time-dependent functions by variational parameters associated with the operators $\hat{G}_j$. In that setting, the ansatz takes the form:
\begin{eqnarray}
    \hat{U}(\boldsymbol{\theta}) = e^{-i\sum_j^\eta \theta_j \hat{G}_j}.
\end{eqnarray}
This approach has also been applied to molecular simulations \cite{FerreiroVelezArXiv2024}. However, the same issue of a rapidly increasing $\eta$ appears for higher $l$-th AGP approximations. In this work, we instead propose using the full set $\{ G_j \}_{j=1}^{\eta}$ as the operator pool, and employing ADAPT-VQE to select the most relevant terms. This strategy enables the use of $l>1$, since increasing $l$ only enlarges the operator pool rather than the circuit depth of the ansatz, so that only the most relevant contributions from higher-order AGP approximations are incorporated into the ansatz.
We refer to this algorithm as CD-ADAPT from here on, and define it as follows:
\begin{algorithm}[H]
\caption{CD-ADAPT}
\label{alg:cd-adapt-vqe}
\begin{algorithmic}[1]
    \Require Final Hamiltonian \(\hat{H}_f\), gradient threshold \(\epsilon\), 
    \State Construct from \(\hat{H}_f\): the initial Hamiltonian \(\hat{H}_i\), the \(l\)-th order AGP operators \(\hat{A}_\lambda^{(l)}(t)\), and the initial state \(\ket{\text{HF}}\)
    \State Generate the operator pool \(\{\hat{G}_j\}_{j=1}^\eta\) from \(\hat{A}_\lambda^{(l)}(t)\)
    \State Initialize reference state: \(\ket{\psi} \gets \ket{\text{HF}}\)
    \State Initialize gradient vector: \(\vec g \gets \vec 0\)
    \State Initialize operator ansatz: \(\text{OpAnsatz} \gets \emptyset\)
    \While{\(\|\vec g\|_2 > \epsilon\)} 
        \State Estimate the gradient vector \(\vec g\):
        \For{\textbf{each} \(G_j\) in operator pool}
            \State \(g_j \gets \bra{\psi}\,[\,\hat{H}_f,\;\hat{G}_j\,]\,\ket{\psi}\)
        \EndFor
        \State \(K \gets \arg\max_j \lvert g_j \rvert\)
        \State Select \(\hat{G}_K\)
        \State Append \(\hat{G}_K\) to the operator ansatz: \(\text{OpAnsatz} \gets \text{OpAnsatz} \cup \{\hat{G}_K\}\)
        \State Execute a VQE sub‐routine optimizing over parameters \(\vec\theta\) with current ansatz, then update  
               \(\displaystyle \ket{\psi} \gets \prod_{j=1}^{\mathrm{len}(\text{OpAnsatz})} e^{\,i\,\theta_j\,\hat{G}_j}\;\ket{\psi}\)
    \EndWhile
    \State \Return \(\ket{\psi}\) and \(E = \bra{\psi}\,\hat{H}_f\,\ket{\psi}\)
\end{algorithmic}
\end{algorithm}

\section{Results}

In this section we present the main results obtained with the CD-ADAPT algorithm. We first examine the size of the operator pool generated by our construction and the effect of the approximations involved. We then report the energies obtained for three molecular systems. Finally, we benchmark the performance of CD-ADAPT against ADAPT-VQE and digitized counterdiabatic quantum algorithms.

\subsection{Operator pool size}
\begin{figure}[H]
    \centering    
    \includegraphics[width=0.99\linewidth]{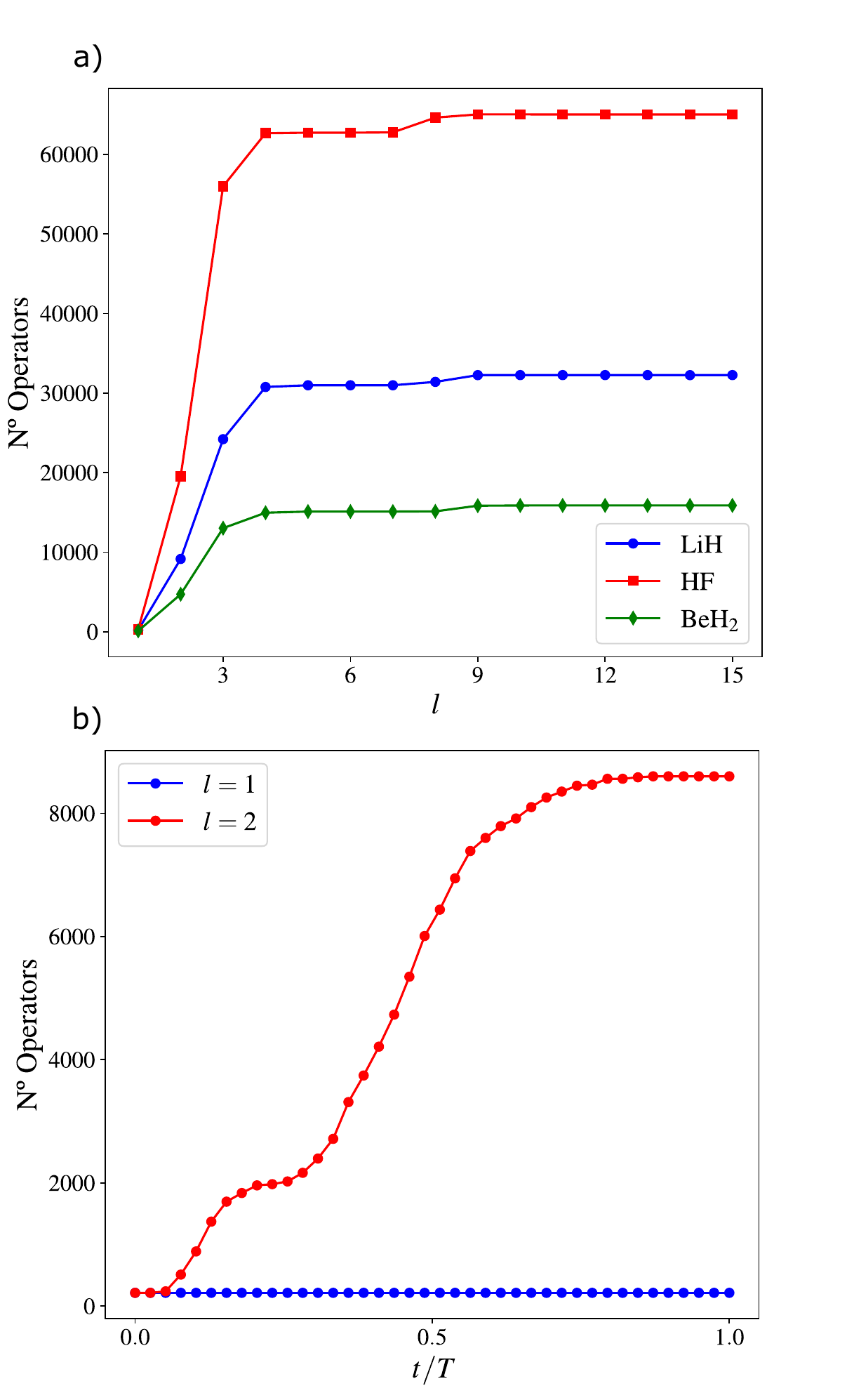}
    \caption{Number of operators in the operator pool $\{ \hat{G}_j \}_{j=1}^{\eta}$ for: a)  LiH, HF and BeH$_2$ in function of $l$-th order of approximation of AGP and: b) for LiH with time-dependent approximation for $l=1,2$.}
    \label{plot1} 
\end{figure}

In our algorithm, the size of the operator pool depends strongly on the specific final Hamiltonian $\hat{H}_f$, since the operators $\{G_j\}_{i=1}^{\eta}$ arise from nested commutators involving both $\hat{H}_i$ and $\hat{H}_f$. Therefore, the election of molecular basis and mapping to qubit operators are determinant in the operator pool in a more relevant way that standard fermionic excitations operator pool since these mappings constrains to a certain subset of the \textit{n}-qubit Pauli group that we will use. An advantage of our algorithm is we can increase the precisión by obtain an more precisely operator pool by chose $l$-th order approximation of AGP. For example for $l=1$ we have:
\begin{eqnarray}
    \hat{A}_\lambda^{l=1} = i\alpha_1(t)[\hat{H}_i,\hat{H}_f],
\end{eqnarray}
and the Pauli strings $\hat{G}_j$:
\begin{eqnarray}
    \hat{G}_j\in \{[\hat{H}_i,\hat{H}_f]\},
\end{eqnarray}
are obtained by computing the commutator using the algebra of the Pauli Group. For $l=2$ the expression is more complex but is possible obtain by simple rules of commutators:
\begin{eqnarray}
    \notag 
\hat{A}_\lambda^{l=2} &=& 
i\alpha_1(t)\,[\hat{H}_i,\hat{H}_f] 
+ i\alpha_2(t)\left((1-\lambda^2(t)[\hat{H}_i,[\hat{H}_i,[\hat{H}_i,\hat{H}_f]]]\right. \\
 \notag 
&+& \lambda(t)(1-\lambda(t))\big([\hat{H}_i,[\hat{H}_f,[\hat{H}_i,\hat{H}_f]]] + [\hat{H}_f,[\hat{H}_i,[\hat{H}_i,\hat{H}_f]]]\big)\\
&+& 
\left. \lambda^2(t) [\hat{H}_f,[\hat{H}_f,[\hat{H}_i,\hat{H}_f]]] \right),
\end{eqnarray}
and the Pauli strings $\hat{G}_j$ are the operators that appear when evaluating these commutators:
\begin{eqnarray}\label{eq:2nd_order_paulistring}
    \notag \hat{G}_j\in \{ [\hat{H}_i,\hat{H}_f],[\hat{H}_i,[\hat{H}_i,[\hat{H}_i,\hat{H}_f]]],
    [\hat{H}_f,[\hat{H}_i,[\hat{H}_i,\hat{H}_f]]] \\
    ,  [\hat{H}_f,[\hat{H}_i,[\hat{H}_i,\hat{H}_f]]], [\hat{H}_f,[\hat{H}_f,[\hat{H}_i,\hat{H}_f]]]\},
\end{eqnarray}
where we observe a substantial increase in the number of operators in the pool for $l=2$. We use the tensorized Pauli decomposition (TPD) algorithm \cite{HantzkoPhysScr2024} to extract the Pauli strings from the approximate AGP. As an illustration, for the molecules lithium hydride ($\text{LiH}$), hydrogen fluoride ($\text{HF}$), and linear beryllium hydride ($\text{BeH}_2$), and using a 10-qubit model, the corresponding sizes of the operator pool are:

\begin{table}[h]
\centering
\begin{tabular}{|c|c|c|}
\hline
  &Nº Operators ($l=1$) & Nº Operators ($l=2$) \\ \hline
LiH & 216 & 9148 \\ \hline
HF & 288 & 19540  \\ \hline
BeH$_2$ & 108 & 4712  \\ \hline
\end{tabular}
\caption{Operator pool size $\eta$ for $l=1$ and $l=2$.}
\end{table}
The growth of $\eta$ from $l=1$ to $l=2$ is substantial, and one might expect a continued increase with higher $l$-th orders. However, the exact AGP provides an upper bound on the number of Pauli strings that can be generated in its approximate AGP. Figure~\ref{plot1}(a) shows numerical results utilizing the  TPD algorithm for the size of the operator pool at different orders $l$ for the LiH, HF and BeH$_2$ molecules, where we observe that the sequence converges towards a constant value. 

It would be desirable to obtain an intermediate number of operators between different $l$-th orders, particularly at low orders where the variation is the most significant. To this end, we propose an approach based on approximations that are induced by the amplitude order of the schedule function $\lambda(t)$, which, combined with the TPD algorithm, yields different numbers of operators depending on the chosen time using the time-dependent Hamiltonian of equation \ref{cd_hamiltonian}. Recall that $\lambda(t)$ is time dependent, with $\lambda(0)=0$ at the initial time, $\lambda(T)=1$ at the final time, and $\dot{\lambda}(0)=\dot{\lambda}(T)=0$. The schedule function used in this work is:
\begin{eqnarray}
    \lambda(t) = \sin^2 \left ( \frac{\pi}{2} \sin^2 \left (\frac{\pi t}{2 T} \right ) \right ). \label{Eq09}
\end{eqnarray}
We use the nested commutator expressions as functions of time, but retaining only the dependence on $\lambda(t)$, without including $\alpha_k(t)$ or $\dot{\lambda}(t)$. This choice is justified because $\dot{\lambda}(t)$ multiplies all commutators uniformly and therefore does not affect the number of operators, while $\alpha_k(t)$ is obtained by minimizing the action $S_l$ and can be absorbed into the variational parameter to be optimized. For example, for $l=2$, the time-dependent function that defines the operator pool is:
\begin{eqnarray}
   \notag \hat{G}_j^{t'} &\in& \left\{[\hat{H}_i,\hat{H}_f], (1-\lambda^2(t'))[\hat{H}_i,[\hat{H}_i,[\hat{H}_i,\hat{H}_f]]]\right. \\
 \notag 
&+&\lambda(t')(1-\lambda(t'))\big([\hat{H}_i,[\hat{H}_f,[\hat{H}_i,\hat{H}_f]]] + [\hat{H}_f,[\hat{H}_i,[\hat{H}_i,\hat{H}_f]]]\big)\\
&+& \left.  \lambda^2(t') [\hat{H}_f,[\hat{H}_f,[\hat{H}_i,\hat{H}_f]]] \right\}, 
\end{eqnarray}
where $t' \in [0,T]$ is the chosen time. For different values of $t'$, the number of operators in the pool changes because the TPD algorithm discards Pauli strings whose coefficients fall below a given threshold, and $\lambda(t')$ modulates these coefficients. It is important to note that this method for operator pool makes use of the TPD algorithm, which relies on the matrix representation of the commutators and is therefore not scalable with the number of qubits. Nevertheless, our algorithm does not depend on TPD algorithm, which is employed here solely for numerical analysis. Other criteria, as proposed in \cite{VanDykePhysRevResearch2024, LongPhysRevA2024}, may also be used to obtain an intermediate operator pool size between $l=1$ and $l=2$ without matrix representation. 

To illustrate this approach, Fig.~\ref{plot1}~b) shows that for $l=2$ and the LiH molecule the number of operators varies as a function of time. This behavior can be exploited for different $l$-th orders, providing a way to obtain intermediate operator pool size between $l=1$ and higher $l$-th orders. We examine how this approach for operator pool influences the performance of the CD-ADAPT algorithm in the following section.

\subsection{Numerical results with CD-ADAPT algorithm}

Numerical simulations were performed using the Qiskit SDK \cite{QiskitArXiv2024} to implement the TPD algorithm for operator pool generation for $l=1$ and $l=2$, and to execute the CD-ADAPT algorithm. The Qiskit Nature library \cite{QiskitNatureZenodo2023} was employed to construct the molecular Hamiltonians $\hat{H}_f$ of the form of Eq. \ref{eq:H_es_Pauli} via the built-in PySCF driver and the mappers provided. The performance of the CD-ADAPT algorithm was evaluated for the $\text{LiH}$, $\text{HF}$, and  $\text{BeH}_2$ molecules. For the $\text{LiH}$ and $\text{BeH}_2$ the electronic structure Hamiltonian was initialized across a range of interatomic distances. We employed an \textit{Active Space Transformation} consisting of 4 electrons and 5 spatial orbitals resulting in encoding 10 spin-orbitals in 10 qubits, using the \text{STO-3G} basis set and the Jordan-Wigner mapping. For the $\text{HF}$ molecule, the same active space parameters were applied, but utilizing the larger \text{6-31G} basis set. Optimization within the VQE subroutines was performed using the L-BFGS-B optimizer, with the gradient convergence threshold set to $\epsilon = 10^{-2}$.

\begin{figure}[H]
    \centering

    \begin{subfigure}
        \centering
        \begin{overpic}[width=0.99\linewidth]{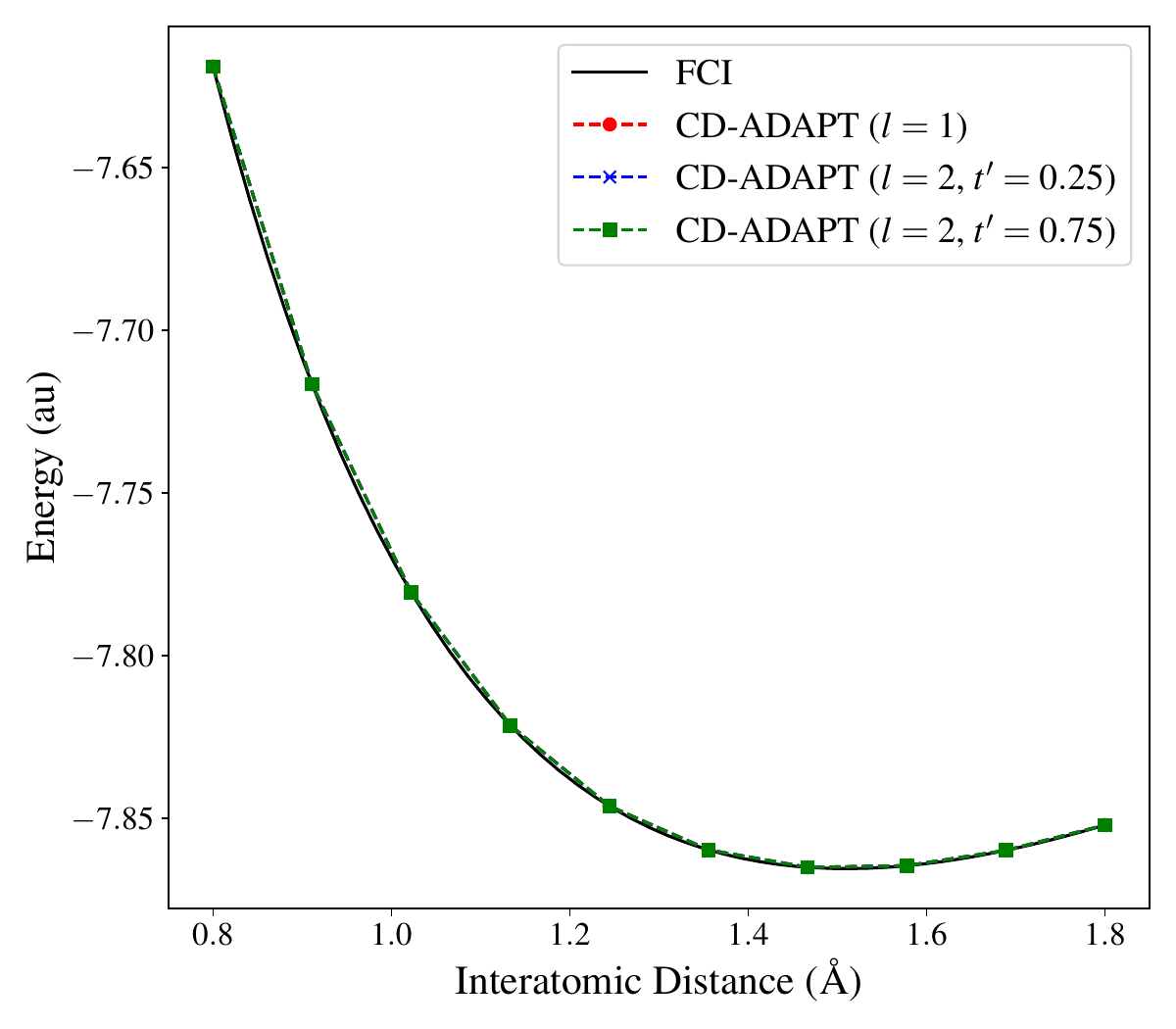}
            \put(2,91){a)}
        \end{overpic}
    \end{subfigure}

    \vspace{0.4em}

    \begin{subfigure}
        \centering
        \begin{overpic}[width=0.99\linewidth]{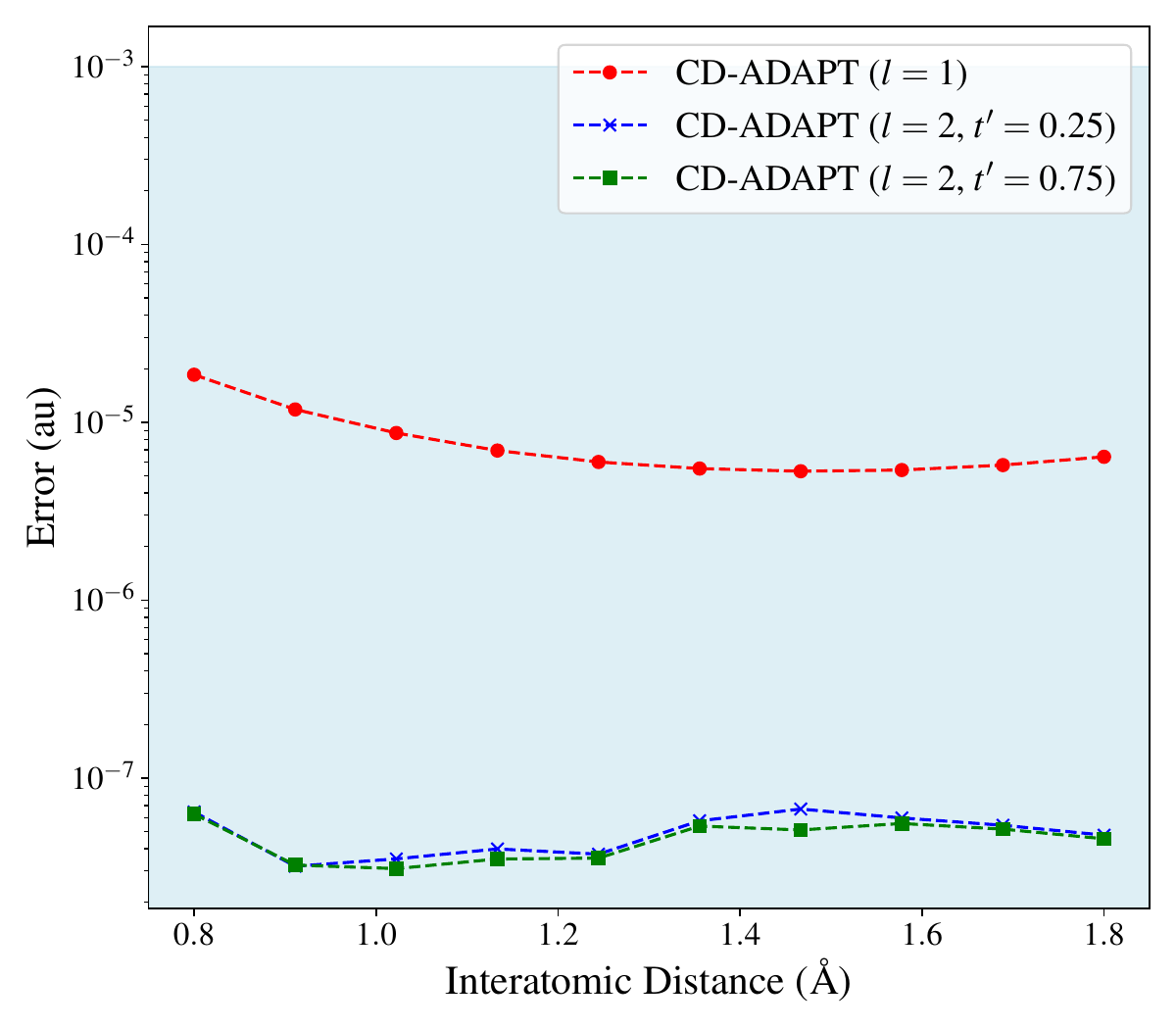}
            \put(2,91){b)}
        \end{overpic}
    \end{subfigure}

    \caption{Ground state energy and corresponding error for LiH calculated via CD-ADAPT. (a) Energy dissociation curve compared to the FCI. (b) Absolute error as a function of interatomic distance for first ($l=1$) and second-order ($l=2$) approximations with time parameters $t'=0.25$ and $t'=0.75$.}
    \label{fig:LiH_results}
\end{figure}

\begin{figure}[H]
    \centering

    \begin{subfigure}
        \centering
        \begin{overpic}[width=0.99\linewidth]{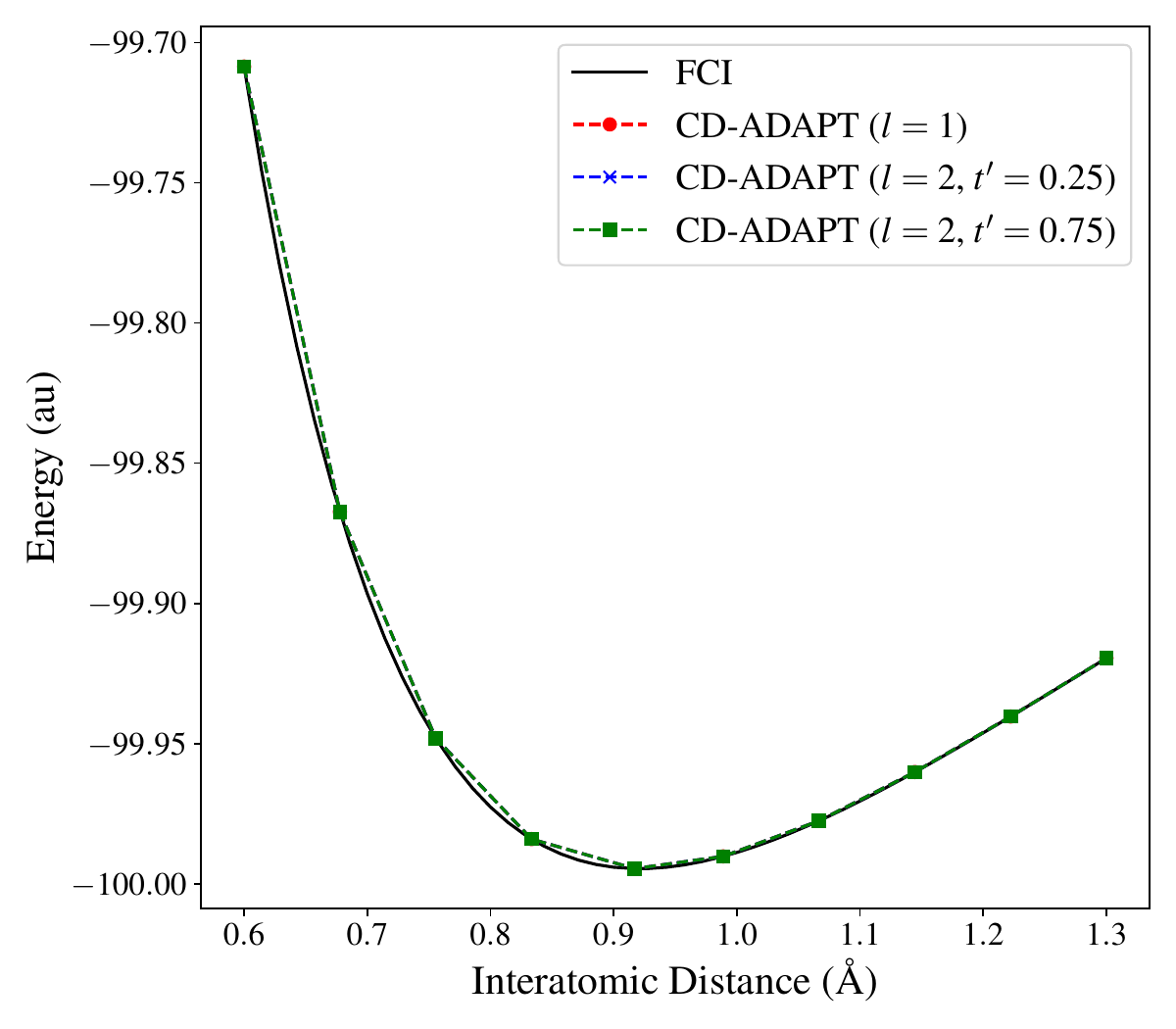}
            \put(2,91){a)}
        \end{overpic}
    \end{subfigure}

    \vspace{0.4em}

    \begin{subfigure}
        \centering
        \begin{overpic}[width=0.99\linewidth]{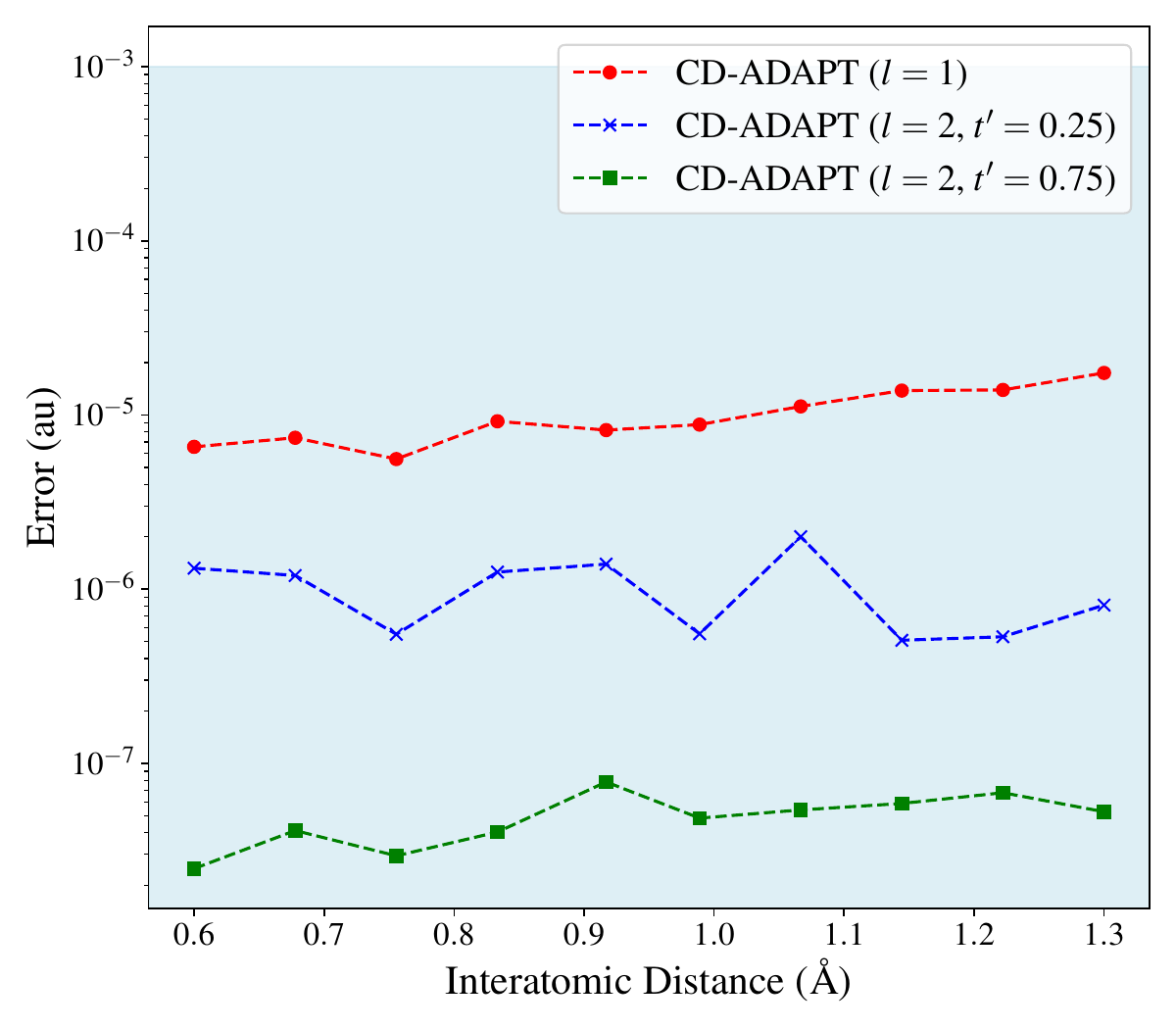}
            \put(2,91){b)}
        \end{overpic}
    \end{subfigure}

    \caption{Ground state energy and corresponding error for HF calculated via CD-ADAPT. (a) Energy dissociation curve compared to the FCI. (b) Absolute error as a function of interatomic distance for first ($l=1$) and second-order ($l=2$) approximations with time parameters $t'=0.25$ and $t'=0.75$.}
    \label{fig:HF_results}
\end{figure}

\begin{figure}[H]
    \centering

    \begin{subfigure}
        \centering
        \begin{overpic}[width=0.99\linewidth]{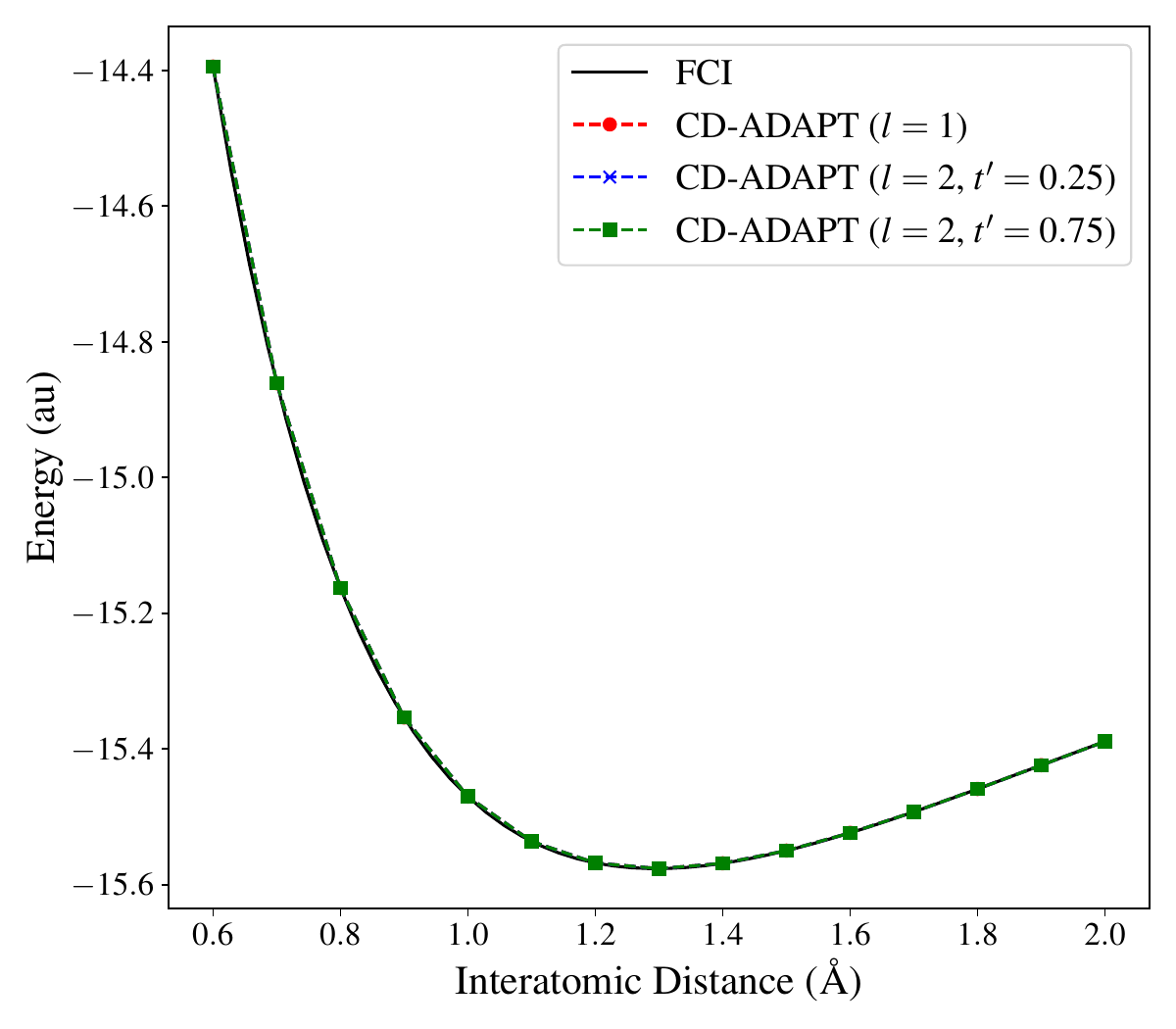}
            \put(2,91){a)}
        \end{overpic}
    \end{subfigure}

    \vspace{0.4em}

    \begin{subfigure}
        \centering
        \begin{overpic}[width=0.99\linewidth]{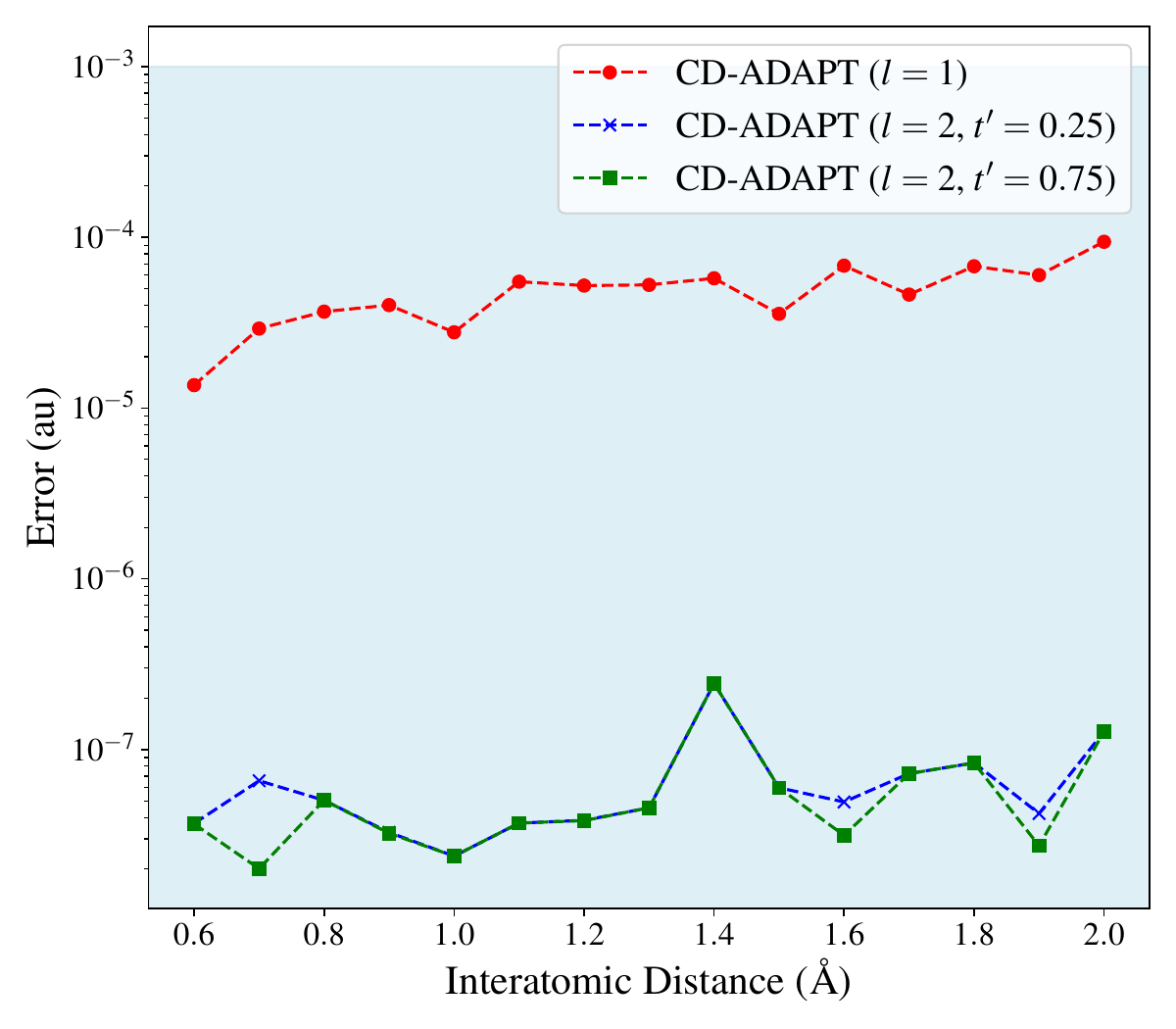}
            \put(2,91){b)}
        \end{overpic}
    \end{subfigure}

    \caption{Ground state energy and corresponding error for BeH$_2$ calculated via CD-ADAPT. (a) Energy dissociation curve compared to the FCI. (b) Absolute error as a function of interatomic distance for first ($l=1$) and second-order ($l=2$) approximations with time parameters $t'=0.25$ and $t'=0.75$.}
    \label{fig:BeH2_results}
\end{figure}

As discussed in the previous section, the number of operators generated by the $l$-th order approximate AGP is inherently dependent on the approximation order $l$ and the specific time $t'$ at which the AGP is evaluated if the TDP algorithm is used to interpolate between different orders of $l$. We conducted a comparison to determine how these parameters impact in the precision in recovering the ground-state energy. Specifically, we analyzed the convergence behavior for $l = \{1, 2\}$ and, for $l=2$, compared time points $t' = \{0.25, 0.75\}$. The results of these simulations are illustrated in Fig. \ref{fig:LiH_results}, \ref{fig:HF_results} and \ref{fig:BeH2_results}.

The performance of the CD-ADAPT ansatz for the LiH molecule was evaluated by analyzing the ground state energy surface and the corresponding absolute errors relative to the Full Configuration Interaction (FCI) benchmark (see Fig.~\ref{fig:LiH_results}). Regarding the expansion order $l$, Fig.~\ref{fig:LiH_results}(a) shows that the first-order approximation ($l=1$) successfully captures the qualitative dissociation profile. The error analysis in Fig.~\ref{fig:LiH_results}(b) reveals that $l=1$ yields errors in the order of $10^{-5}$~Ha. Increasing the order to $l=2$ results in a significant accuracy improvement, reducing errors by approximately up to two orders of magnitude (down to $10^{-5}$--$10^{-7}$~Ha), highlighting that generating the operator pool from higher-order counterdiabatic terms yields more expressive operators. Furthermore, the results demonstrate the influence of the evolution time $t'$. For the second-order approximation ($l=2$), the $t'=0.75$ consistently yields lower errors compared to the pool generated with the approximated AGP potential evaluated at $t'=0.25$.\\

The CD-ADAPT algorithm was further tested on the HF molecule, a system exhibiting significant electron correlation effects driven by the strong polarity induced by the fluorine atom. The computed energy landscape and the corresponding absolute error profiles relative to the exact FCI benchmark are presented in Fig.~\ref{fig:HF_results}. Regarding the expansion order, the data in Fig.~\ref{fig:HF_results}(a) confirms that $l=1$ is sufficient to reproduce the behavior of the dissociation curve under chemical accuracy, yielding errors consistently around $10^{-5}$~Ha (see Fig.~\ref{fig:HF_results}(b)). Conversely, the inclusion of $l=2$ in the operator pool yields a drastic reduction in the error. Notably, the error drops by several orders of magnitude, reaching the $10^{-6}$--$10^{-8}$~Ha. This behavior reinforces the notion that the operator pool derived from higher-order nested commutators is more effective in expressing unitary operators that approximates the HF ground state. A comparison within $l=2$ reveals that $t' = 0.75$ significantly outperforms $t' = 0.25$. As shown in the logarithmic error plot, operator pool with $t' = 0.75$ allows the ansatz to suppress deviations down to $10^{-7}$~Ha near the equilibrium geometry.

Finally, we tested our algorithm on the BeH$_2$ molecule. Consistent with the previous results, as shown in Fig. \ref{fig:BeH2_results}, using the operator pool with $l=1$ yields errors below chemical accuracy and on the order of $10^{-5}\,\text{Ha}$. In this case, the $l=2$ operator pool exhibits significantly better performance than $l=1$, while showing similar performance for $t' = 0.25$ and $t' = 0.75$ across all considered bond distances, with errors on the order of $10^{-7}$--$10^{-8}\,\text{Ha}$. This shows that using more operators from the $l=2$ approximation does not always guarantee better accuracy, and that employing a reduced subset of these operators can already lead to a significant improvement compared to the $l=1$ case. Future work may focus on developing improved criteria for selecting operators from the $l=2$ or higher-order approximations of the AGP, with the aim of constructing a more compact operator pool. 

\subsection{Comparison with digitized counterdiabatic quantum optimization and fermionic ADAPT-VQE}


To benchmark the performance of the proposed CD-ADAPT algorithm \ref{alg:cd-adapt-vqe}, we compared it against the two methods that inspired its development: the DCQO and the fermionic ADAPT-VQE. The DCQO was implemented using Qiskit by performing a Trotterization of the unitary evolution operator defined in Eq. \ref{eq:trotterized_U(T)_at_final_time}, with a fixed total evolution time $T=1$ and $N_t = 2$ Trotter steps. The fermionic ADAPT-VQE was implemented utilizing the built-in solvers provided by Qiskit Nature \cite{QiskitNatureZenodo2023} and Qiskit Algorithms.

\begin{figure}[H]
    \centering

    \begin{subfigure}
        \centering
        \begin{overpic}[width=0.99\linewidth]{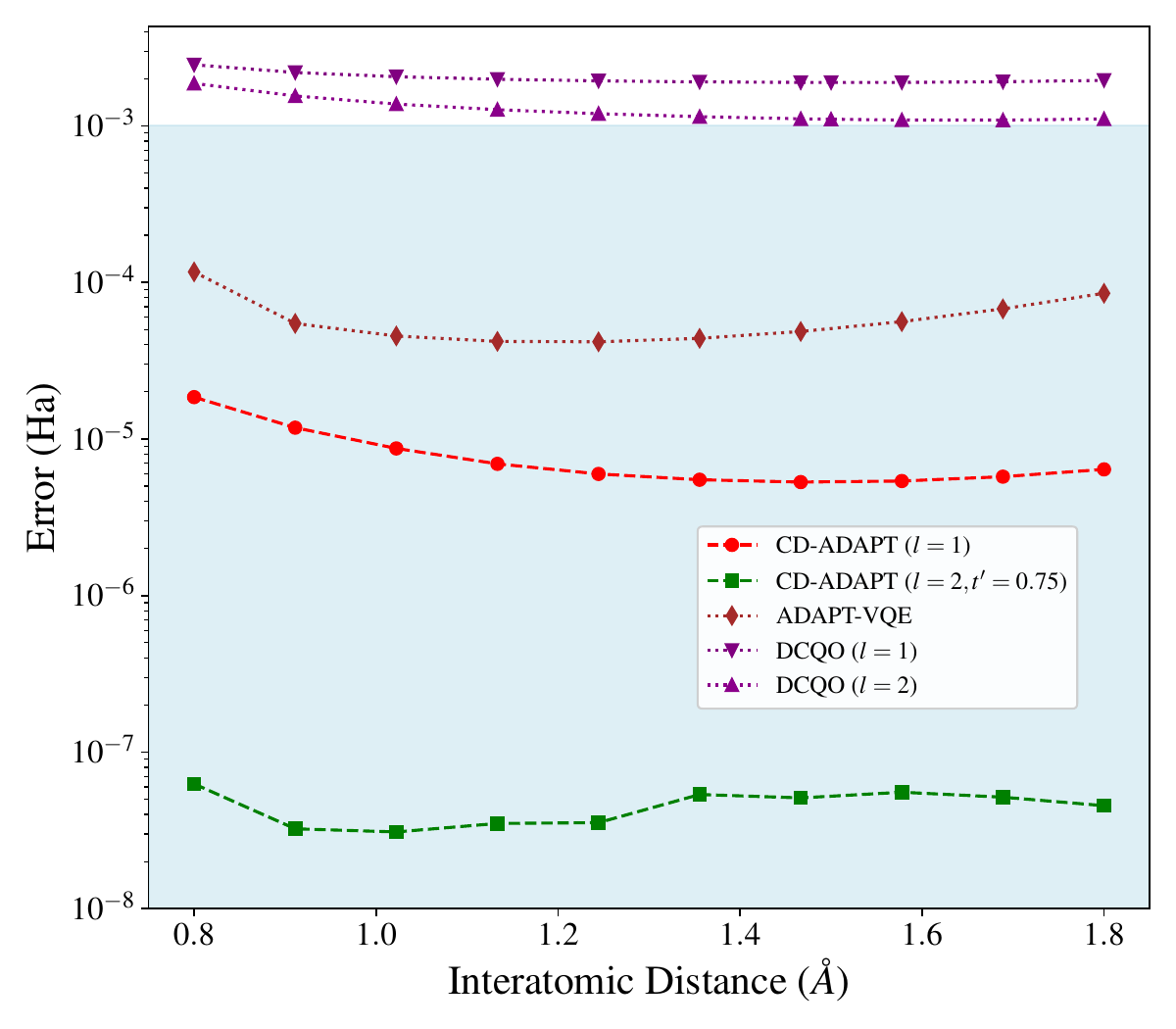}
            \put(2,91){a)}
        \end{overpic}
    \end{subfigure}

    \vspace{0.4em}

    \begin{subfigure}
        \centering
        \begin{overpic}[width=0.99\linewidth]{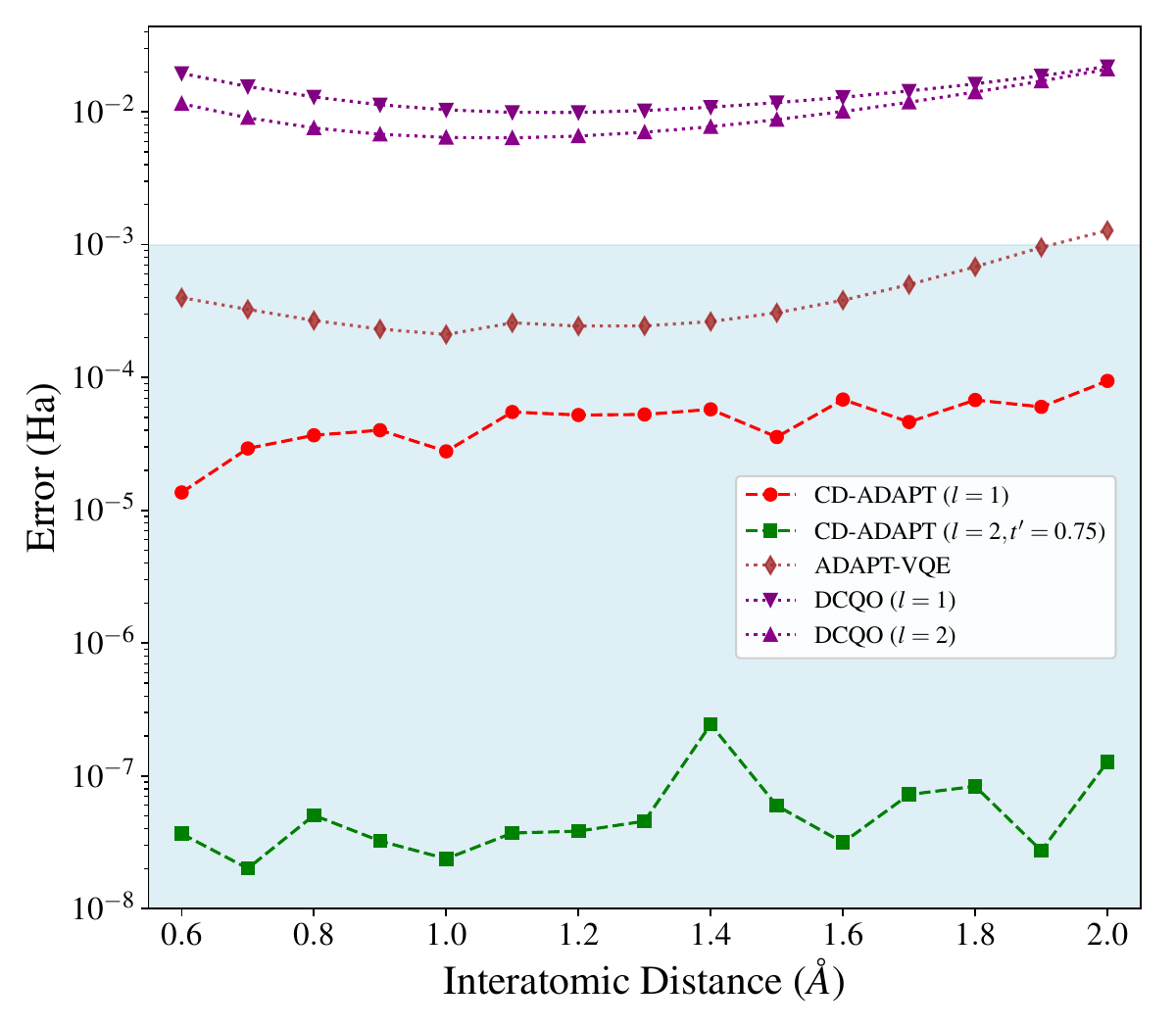}
            \put(2,91){b)}
        \end{overpic}
    \end{subfigure}

    \caption{Absolute error calculated for CD-ADAPT, ADAPT-VQE and DCQO vs interatomic distance. (a) For LiH (b) For BeH$_2$.}
    \label{fig:Comparison_LiH_BeH2}
\end{figure}

We analyzed the ground state accuracy relative to the FCI benchmark achieved by these algorithms across various interatomic distance intervals. Figure~\ref{fig:Comparison_LiH_BeH2} shows that CD-ADAPT consistently achieves higher accuracy than the other algorithms considered, more precisely, with respect to the ADAPT-VQE algorithm, the $l=1$ instance of the CD-ADAPT achieves one order of magnitude improvement in the estimation of the ground state energy, while for the $l=2$ instance and $t'=0.75$ shows an improvement of three orders of magnitude. With respect to the DCQO, this improvements scales up to three and five orders of magnitude, respectively. We also observe that the DCQO remains above chemical accuracy for $N_t = 2$, indicating a poor approximation of the counterdiabatic evolution with the selected numerical parameters. This value of $N_t$ was intentionally chosen to analyze the number of controlled-NOT (CNOT) gates, revealing that DCQO is already less efficient than our approach in terms of circuit complexity. Although increasing $N_t$ would improve the accuracy, it is done at the cost of a significantly larger number of CNOT, rendering the method impractical. Furthermore, we evaluate the circuit complexity by quantifying the number of CNOT gates required to implement each ansatz at a fixed interatomic distance for each molecule. Circuit transpilation was performed using the \texttt{transpile} subroutine in Qiskit with optimization level 3 on the \texttt{GenericBackendV2}. The comparative results are detailed in Table \ref{tab:BeH2comparison} for $\text{BeH}_2$ ($r= 1.50$ \AA), Table \ref{tab:LiHcomparison} for $\text{LiH}$ ($r= 1.58$ \AA), and Table \ref{tab:HFcomparison} for $\text{HF}$ ($r = 0.917$ \AA).

\begin{table}[H]
\centering
\begin{tabular}{|c| c| c| c|}
\hline
Algorithm & Error (au) & N° parameters & N° CNOTs \\
\hline
CD-ADAPT ($l=1$)  & $3.56 \cdot 10^{-5}$  & 18  & 208  \\
CD-ADAPT ($l=2$, $t'=0.25$)  & $5.98 \cdot 10^{-8}$ & 27  & 318  \\
CD-ADAPT ($l=2$, $t'=0.75$) & $5.98 \cdot 10^{-8}$ & 27 & 324 \\
ADAPT-VQE & $ 3.07 \cdot 10^{-4}$ & 6 & 419 \\
DCQO ($l=1$) & $1.17 \cdot 10^{-2}$ & - & 1359 \\
DCQO ($l=2$) & $8.72 \cdot 10^{-3}$ & - & 5665 \\
\hline
\end{tabular}
\caption{Comparison of the error, number of variational parameters, and CNOT gates for different algorithms applied to $\text{BeH}_2$ at distance $r=1.500  \AA $. For all ADAPT-based algorithms, the threshold is fixed at $\epsilon = 10^{-2}$, while for DCQO the number of Trotter steps is set to two. }
\label{tab:BeH2comparison}
\end{table}


\begin{table}[H]
\centering
\begin{tabular}{|c| c| c| c|}
\hline
Algorithm & Error (au) & N° parameters & N° CNOTs \\
\hline
CD-ADAPT ($l=1$)  & $5.39 \cdot 10^{-6}$  & 12 & 134  \\
CD-ADAPT ($l=2$, $t'=0.25$)  & $5.96 \cdot 10^{-8}$ & 32 & 368 \\
CD-ADAPT ($l=2$, $t'=0.75$) & $5.55 \cdot 10^{-8}$ & 31 & 376 \\
ADAPT-VQE & $5.61 \cdot 10^{-5}$ & 10 & 884 \\
DCQO ($l=1$) & $1.89 \cdot 10^{-3}$ & - & 1415 \\
DCQO ($l=2$) & $1.09 \cdot 10^{-3}$ & - & 5269 \\
\hline
\end{tabular}
\caption{Comparison of the error, number of variational parameters, and CNOT gates for different algorithms applied to $\text{LiH}$ at distance $r=1.578 \AA $. For all ADAPT-based algorithms, the threshold is fixed at $\epsilon = 10^{-2}$, while for DCQO the number of Trotter steps is set to two. }
\label{tab:LiHcomparison}
\end{table}

\begin{table}[H]
\centering
\begin{tabular}{|c| c| c| c|}
\hline
Algorithm & Error (au) & N° parameters & N° CNOTs \\
\hline
CD-ADAPT ($l=1$)  & $8.19 \cdot 10^{-6}$ & 14 & 156\\
CD-ADAPT ($l=2$, $t'=0.25$)  & $1.39 \cdot 10^{-6}$ & 25 & 274 \\
CD-ADAPT ($l=2$, $t'=0.75$) & $7.82 \cdot 10^{-8}$ & 36 & 408 \\
ADAPT-VQE & $1.82 \cdot 10^{-5}$ & 12 & 993 \\
DCQO ($l=1$) & $2.66 \cdot 10^{-2}$ & - & 2343 \\
DCQO ($l=2$) & $5.58 \cdot 10^{-3}
$ & - & 14083 \\
\hline
\end{tabular}
\caption{Comparison of the error, number of variational parameters, and CNOT gates for different algorithms applied to $\text{HF}$ at distance $r= 0.917 \AA $. For all ADAPT-based algorithms, the threshold is fixed at $\epsilon = 10^{-2}$, while for DCQO the number of Trotter steps is set to two. }
\label{tab:HFcomparison}
\end{table}
The results demonstrate that, for the given gradient threshold, our CD-ADAPT approach significantly outperforms the benchmark methods in terms of both accuracy and circuit efficiency (CNOT count). As shown at Tables \ref{tab:BeH2comparison}, \ref{tab:LiHcomparison} and \ref{tab:HFcomparison}, the CD-ADAPT algorithm requires $\sim 100-500$ fewer CNOT gates than ADAPT-VQE for this specific configuration. Although ADAPT-VQE generally employs fewer variational parameters, this comes at the cost of lower accuracy for the same convergence threshold. In contrast, CD-ADAPT constructs a more extensive pool of operators derived from the counterdiabatic expansion, enabling deeper convergence with more compact circuits.

\section{Conclusion} 
We proposed a hybrid quantum–classical algorithm that constructs an operator pool from  the counterdiabatic protocol using aproximate adiabatic gauge potential and combines it with the ADAPT-VQE \cite{GrimsleyNatCommun2019}  strategy to build the ansatz. This approach enabled the inclusion of higher order of approximate adiabatic gauge potential without a corresponding increase in the ansatz circuit depth as observed in conventional DCQO \cite{HegadePRResearch2022}, since the ADAPT-VQE gradient criterion selects only the more relevants operators for approaching to ground state. Numerical simulations show that our algorithm achieves a lower number of CNOT gates and an error reduced by approximately five orders of magnitude compared to DCQO with a Trotter step number equal to two. On the other hand, the operator pool obtained from the counterdiabatic protocol appeared to be more efficient than that constructed from fermionic singles and doubles excitations. This becomes evident when comparing with the fermionic ADAPT-VQE algorithm, where our approach achieves a lower number of CNOT gates and an error that is three orders of magnitude smaller. These results indicated that accelerated adiabatic dynamics assisted by counterdiabatic driving can provide a more effective route to the ground state of molecular Hamiltonian than operator selections based solely on physically motivated interaction terms, such as single and double excitations. 

To follow the counterdiabatic protocol, we proposed an initial Hamiltonian whose ground state is the Hartree--Fock state and which is composed of local operators. This choice requires prior knowledge of the Hartree--Fock state, which is typically known and used as the reference state in molecular simulation algorithms. In addition, we employed the TPD algorithm to obtain the Pauli strings associated with the approximate adiabatic gauge potential at orders $l=1$ and $l=2$. For $l=2$, a fixed time dependence was considered, where the selected times are $t' = 0.25$ and $t' = 0.75$, allowing us to generate operator pools with intermediate sizes between the $l=1$ and $l=2$. The numerical simulations of our algorithm were performed for the $\text{LiH}$, $\text{HF}$, and $\text{BeH}_2$ molecules, yielding results below chemical accuracy when using the operator pool with $l=1$, and achieving an improvement of approximately two orders of magnitude when increasing the operator pool to $l=2$. These results demonstrated a suitable performance for NISQ and early fault-tolerant quantum computing in molecular simulation.

\section{Data Availability}
The data that support the findings of this study are available from the corresponding author upon reasonable request.

\section{Acknowledgement}
DT acknowledge grant Posdoctorado UC PD2024-609. HD and DG acknowledge grant FONDECyT Regular nr 1230586, Chile. HD acknowledge ANID BECAS/MAGÍSTER NACIONAL 22251911

\end{document}